\documentstyle[twoside,fleqn,espcrc2,epsf]{article}

\newcommand{\AmS}{{\protect\the\textfont2
  A\kern-.1667em\lower.5ex\hbox{M}\kern-.125emS}}

\title{Nonperturbative ``Lattice Perturbation Theory''\thanks{Talk given
by Paul Mackenzie  }
}
\author{W. Dimm,\address{Floyd R. Newman Laboratory of Nuclear Studies, Cornell University, Ithaca, NY 14853 USA}
 G. Peter Lepage,$^a$ 
 and Paul B. Mackenzie,\address{Theoretical Physics Dept., Fermilab, P.O. Box 500, Batavia, IL  60510  USA}
}
       
\begin{document}

\begin{abstract}
 We discuss a program for replacing 
standard perturbative methods with
Monte Carlo simulations in
 short distance lattice gauge theory
calculations.	
\end{abstract}

\maketitle

In principle, perturbation theory is unnecessary to solve QCD with lattice 
methods.  Even the short distance calculations relating the lattice
action to  continuum actions
could in principle be done by $a\rightarrow 0, V\rightarrow \infty$
brute force Monte Carlo calculations.
In practise, perturbation theory has been essential to the progress 
of lattice methods.
It is much easier and more powerful than Monte Carlo for some
purposes:
\begin{itemize}
\item The $a\rightarrow 0, V\rightarrow \infty$ limits are much easier to take
  in perturbative calculations.
\item The values of perturbative coefficients can be calculated much more 
  accurately than typical quantities in numerical calculations.
\item When short distance quantities can be computed with very different 
  calculational methods, such as perturbation theory and Monte Carlo 
  simulation,
  confidence is bolstered in both methods.
\end{itemize}

In this talk, we will reexamine the question of when it may be 
advantageous and feasible to do ``Feynman diagrams''
nonperturbatively.
An important motivation for reexamining methods for perturbative 
calculations now is the increasing understanding and use 
of improved actions.
$O(a)$ improved actions for Wilson fermions are known to be crucial for 
calculating some quantities (such as the hyperfine splitting in 
quarkonium systems).
They may be important for many more quantities.
The perturbation theory for $O(a)$ improved actions is much harder
than the perturbation theory for unimproved actions (already
hard enough), but still tractable.
Most of the most important calculations for this action are in the process
of being done.

The frontier in the practical application of improved action is 
$O(a^2)$ improvement.  For example, in Lepage's talk at this conference,
he showed that a mean-field improved Weisz action\cite{Wei}
(the pure gauge action constructed from plaquettes and flat six-link
loops)
plus the improved fermion action of NRQCD\cite{lep}
is capable of calculating the charmonium spectrum correctly to a few per
cent, even at lattice spacings as large as $a^{-1} \sim$ 400 MeV.
Even if the program to perform spectrum calculations with improved actions
is very successful, that is only part of the total program for
lattice QCD.
Extraction of decay constants, form factors and quark masses 
requires short distance calculations which have been done
with perturbative methods.
The action and Feynman rules for this action are much, much more 
complicated than those for the unimproved theory.
Redoing all existing perturbative calculations with this action
will be terribly complicated,
even with large increases in the amount
of work devoted to perturbative calculations.
The job would be more difficult than the job of redoing all of the
simulation programs for the new action.

An even more extreme example is the ``perfect action''.\cite{Has}
Classical corrections to the action are much more tractable than
quantum corrections.
They can be done more or less ``perfectly'',
but at the cost of adding many additional terms to the action.
The work required for  deriving Feynman rules and performing
perturbative calculations with such an action 
 is hard to imagine.

The purpose of this talk is to ask whether it is possible to perform short 
distance calculations without deriving Feynman rules
and doing Dirac and Lorentz algebra.
There are several possible goals of short distance Monte Carlo
calculations.
One, which is not the subject of this talk but which is potentially more
 important, is the search for nonperturbative effects at short distances.
For example, the expectation value of the plaquette is expected to have,
in addition to its expansion in powers of $g^2$, nonperturbative effects
 which fall off as some power of the lattice spacing.
The condensate picture of short distance nonperturbative effects 
suggests that this power is four, the dimension of the operator
$F_{\mu\nu}^2$. 
Little is known from first principles, however.
This subject is 1) hugely important, and 2) not the subject of this talk.

The questions addressed by  this talk are more modest.
What are the coefficients of the powers of $g^2$ in the perturbative
expansion?
To what extent can we find methods to test perturbation theory where
we can use conventional methods, and to replace perturbation theory where
we cannot?
In cases where the correct perturbative coefficient is known, can we use 
nonperturbative methods to
\begin{itemize}
\item recalculate the first order coefficient correctly?
\item bound or estimate the $O(\alpha^2), O(\alpha^3),  \ldots $
  coefficients correctly?
\item redo the calculation with ten times as complicated an action?
\end{itemize}

Some short distance nonperturbative calculations are easier than others.
For example, much may be learned about the extraction of $\alpha_s$
from nonperturbatively calculated short distance quantities with
relatively simple calculations.\cite{LM}
Other quantities, such as operator normalizations and quark mass extractions,
are more complicated.
As an illustrative example, we will consider one of the more difficult
quantities, extracting the quark mass from nonperturbative short distance
calculations.
The goal will be to determine numerically the 
coefficients of the first power or two in $\alpha_s$ 
 for a given value of the quark mass in 
lattice units, with the idea of using the coefficients
in the way that normal perturbative results
 for the same value of $ma$
would be used 
in phenomenological calculations.
(Perturbative coefficients are explicit functions of the quark mass
when one is not in the limit $a m \rightarrow 0$.)
We will do this by performing a Monte Carlo simulation with very small
lattice spacing, fixing the gauge, creating a quark propagator with wall
source and sink, and calculating an effective (pole) mass
for the quark in the usual way.

There are several  problems which arise in this calculation.
They are mostly associated in one way or another with the limiting procedure.
The nice limit for perturbative calculations is the limit
$V\rightarrow \infty$, followed by $g^2\rightarrow 0$.
But Monte Carlo calculations (on noninfinite computers) require
$V$ finite, while we would still like $g^2 \rightarrow 0$.
This is a complicated limit for the theory.
In particular, as the quark mass in lattice units becomes small,
we do not have the desirable property
$a m >> 1/L$ (where $a$ is the lattice spacing, $m$ is the quark mass, and
$L$ is the box size in lattice units).

One problem is the effect of tunneling between the (81) Z3 vacua of the
gauge theory.  Quarks have energies of order $1/L$ in nontrivial vacua.
Only in the trivial vacuum do these not dominate the $O(g^2)$ correction
unless $L\rightarrow \infty$.
Only the trivial vacuum part of the path integral corresponds to the usual
perturbative momentum sum.

Even in the trivial vacuum, zero modes give a contribution to the quark mass
which vanishes only in the large volume limit.
Dimensional analysis suggests that this effect also goes like $1/L$.
These effects must be calculated.  They are especially important for light 
quarks.

Gauge dependence is an additional problem. 
The pole mass is gauge invariant, but infrared sensitive gauges may have 
problems.
Axial gauges are famously IR sensitive.
Landau gauge (and Feynman and other covariant gauges) have gauge dependent
infrared logarithms in their wave function renormalizations.
They therefore have no isolated poles, but have branchcuts instead.
Among the commonly used gauges in Monte Carlo simulations, only Coulomb
gauges (where an initial gauge fixing to $A_0=0$ gauge fixes the time
direction gauge freedom)
has no known infrared problems.
An additional infrared healthy gauge (perhaps Yennie gauge) would be desirable
in our calculations to test gauge dependence.

\begin{figure}
\epsfxsize=3in
\epsfbox{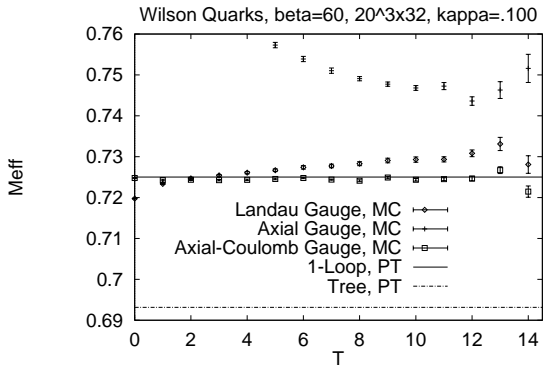}
\vspace{-.4in}
\caption{
Effective mass plot for Wilson quarks in three different gauges
compared with tree level and first order perturbation theory.
}
 \label{figwil}
\end{figure}

\begin{figure}[t]
\epsfxsize=3in
\epsfbox{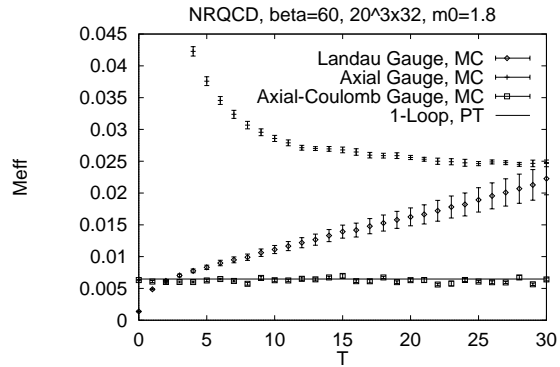}
\vspace{-.4in}
\caption{Effective mass plot for nonrelativistic
 quarks in three different gauges
compared with first order perturbation theory.
}
 \label{fignrq}
\end{figure}

Bearing in mind these complications, which are not all understood and
which may limit the applicability of the approach, we will now ask if a 
Monte Carlo determination of the quark mass correction if feasible.
We perform a simulation with periodic quark boundary conditions (N.B.), 
at $\beta=60.0$,
and at large volume, $20^3\times 32$,  to limit tunneling.
Fig.~\ref{figwil} shows an effective mass plot for Wilson fermions calculated
by Monte Carlo methods in the three gauges mentioned above, and compared with
tree-level and one-loop perturbation theory.  
The gauges with infrared problems (axial and Landau) do not agree well with
one-loop perturbation theory, but the Coulomb gauge results do agree.
Fig.~\ref{fignrq} shows that the same is true for nonrelativistic fermions.

The next practical calculations to be addressed include the calculation 
of the second order additive mass renormalizations for NRQCD and
Wilson fermions, and the first order additive mass renormalizations
for $O(a^2)$ improved actions.
Questions of principle which remain to be fully addressed include
the effects of boundary conditions,  gauge dependence, and
zero modes.
     
Summary:
\begin{itemize}
\item Conventional perturbative calculations get harder faster than Monte
  Carlo calculations as the action gets more complicated.
\item $a^2$ corrected actions will be terribly complicated for perturbation
  theory.
\item Brute force evaluation of quark-gluon Green's functions is clearly
  possible in principle.
\item  Brute force evaluation of quark-gluon Green's functions 
 may also be of practical importance if various complications can be understood.
\end{itemize}

{ ACKNOWLEDGMENTS. \ }
We thank Pierre van Baal,
Aida El-Khadra,
Bart Mertens,
and Andreas Kronfeld
for discussions.
	Fermilab is operated by 	Universities Research
Association, Inc. under contract with the U.S.  	Department of
Energy.


\begin{thebibliography}{9}
\bibitem{Wei}P. Weisz, Nuc. Phys. B212 (1983) 1.
\bibitem{lep}M. Alford, W. Dimm, G. P. Lepage, G. Hockney, and P. B. Mackenzie,
  in these proceedings.
\bibitem{Has}A. Hasenfratz, in these proceedings.
\bibitem{LM} G.P.Lepage and P.B.Mackenzie, Phys. Rev. D48 (1993) 2250.
\end{thebibliography}
\end{document}